\documentclass[conference]{IEEEtran}
\IEEEoverridecommandlockouts
\usepackage{cite}
\usepackage{amsmath,amssymb,amsfonts}
\usepackage{algorithm}
\usepackage{algpseudocode}
\usepackage{graphicx}
\usepackage{textcomp}
\usepackage{xcolor}
\usepackage{subcaption} 
\usepackage{booktabs}
\usepackage{tabularx}
\usepackage{siunitx}
\usepackage{comment}

\usepackage{pifont}
\newcommand{\cone}{\ding{182}}
\newcommand{\ctwo}{\ding{183}}
\newcommand{\cthree}{\ding{184}}
\newcommand{\cfour}{\ding{185}}
\newcommand{\cfive}{\ding{186}}

\usepackage[a4paper, total={184mm,239mm}]{geometry}
\def\BibTeX{{\rm B\kern-.05em{\sc i\kern-.025em b}\kern-.08em
    T\kern-.1667em\lower.7ex\hbox{E}\kern-.125emX}}

\begin{document}

\algnewcommand\algorithmicprocess{\textbf{Process}}
\algdef{SE}[PROCESS]{Process}{EndProcess}[2]{\algorithmicprocess\ #1(#2)}{\algorithmicend\ \algorithmicprocess}

\algdef{SE}[FOR]{pFor}{EndpFor}[1]{\textbf{parallel for} #1 \textbf{do}}{\textbf{end parallel for}}

\newcommand{\qcomm}{{\textsc{qcomm}}}  

\title{Instruction-Directed MAC for Efficient Classical Communication in Scalable Multi-Chip \\Quantum Systems
\thanks{The authors gratefully acknowledge funding from the European Commission through HORIZON-EIC-2022-PATHFINDEROPEN01-101099697 (QUADRATURE) and from the European Union’s Horizon Europe research and innovation programme under grant agreement No. 101042080 (ERC Starting Grant 2021). This work is part of the I+D+i project titled BLOSSOMS, grant PID2024-158530OB-I00, funded by MICIU/AEI/10.13039/501100011033/ and by ERDF/EU.}}

\author{\IEEEauthorblockN{Maurizio Palesi, Enrico Russo, \\Hamaad Rafique, Giuseppe Ascia, \\Davide Patti}
\IEEEauthorblockA{\textit{University of Catania}\\
Catania, Italy \\
firstname.lastname@unict.it}
\and
\IEEEauthorblockN{Abhijit Das}
\IEEEauthorblockA{\textit{Indian Institute of Technology Hyderabad}\\
Telangana, India \\
abhijit.das@cse.iith.ac.in}
\and
\IEEEauthorblockN{Sergi Abadal}
\IEEEauthorblockA{\textit{Universitat Politècnica de Catalunya}\\
Barcelona, Spain \\
abadal@ac.upc.edu}
}

\maketitle

\begin{abstract}
Scalable quantum computing requires modular multi-chip architectures integrating multiple quantum cores interconnected through quantum-coherent and classical links. The classical communication subsystem is critical for coordinating distributed control operations and supporting quantum protocols such as teleportation. In this work, we consider a realization based on a wireless network-on-chip for implementing classical communication within cryogenic environments. Traditional token-based medium access control (MAC) protocols, however, incur latency penalties due to inefficient token circulation among inactive nodes. We propose the instruction-directed token MAC (ID-MAC), a protocol that leverages the deterministic nature of quantum circuit execution to predefine transmission schedules at compile time. By embedding instruction-level information into the MAC layer, ID-MAC restricts token circulation to active transmitters, thereby improving channel utilization and reducing communication latency.
Simulations show that ID-MAC reduces classical communication time by up to 70\% and total execution time by up to 30–70\%, while also extending effective system coherence. These results highlight ID-MAC as a scalable and efficient MAC solution for future multi-chip quantum architectures.
\end{abstract}


\section{Introduction}
The pursuit of large-scale quantum computing is driven by the promise of solving problems intractable for classical machines, from quantum chemistry and materials discovery to secure communication and optimization. While quantum advantage has been experimentally demonstrated on noisy intermediate-scale quantum (NISQ) devices~\cite{arute_nature19}, these systems, typically comprising only tens to a few hundreds of qubits, fall short of the requirements to tackle real-world applications. To move beyond the NISQ era, the community recognizes the necessity of scaling quantum processors to thousands or even millions of qubits. Achieving such scale, however, is fundamentally constrained by technological, architectural, and system-level challenges.

Early efforts toward scalability largely focused on monolithic integration, where qubits and their control electronics are densely packed within a single chip. Although this approach has proven useful for prototyping, it faces severe limitations: wiring congestion, reduced fabrication yield, increased crosstalk, and limited qubit addressability~\cite{almudever_date17,kjaergaard_arcmp20}. These bottlenecks suggest that simply enlarging monolithic chips is not a viable path to scalability. Instead, a paradigm shift is needed towards distributed designs that modularize computation while still preserving quantum coherence across the system.

A promising architectural direction is the multi-chip quantum architecture~\cite{jnane_pra22}, in which the quantum processor is partitioned into multiple moderately sized quantum cores (QCs). Each QC integrates a manageable number of qubits alongside its local control electronics, thereby alleviating wiring density and fabrication yield issues. Crucially, the QCs are interconnected through quantum-coherent links, enabling them to operate as a single logical processor whose computational power scales with the aggregate number of entangled qubits. To orchestrate computation and control across these distributed cores, a complementary interconnection fabric is required, capable of transporting both quantum states and classical control data within the cryogenic package.

This shift towards modular, multi-chip architectures is motivated by the same trajectory that classical computing followed decades ago: from monolithic processors to multi-core and eventually distributed architectures to sustain scaling. For quantum computing, such a transition is not merely a matter of performance but a fundamental requirement to overcome physical constraints at cryogenic temperatures and to enable reliable operation at a large scale. Recent proposals have thus explored hybrid interconnection fabrics that combine quantum-coherent communication channels with cryogenic-compatible wireless networks for classical control~\cite{alarcorn_iscas23}. These developments open the way to a new class of scalable, reconfigurable, and full-stack quantum architectures capable of bridging the gap between near-term devices and large-scale fault-tolerant quantum computers.

Within such multi-chip architectures, the classical communication system plays a pivotal role~\cite{palesi_mcsoc24,escofet_qce25}. On one side, it coordinates system-wide operations through control messages, synchronization, and resource management. On the other, it supports quantum-specific functions such as teleportation, where classical information must accompany quantum state transfer to complete the protocol. In this work, we focus on the classical communication subsystem of scalable multi-chip quantum architectures enabled by cryogenic hybrid wireless/quantum-coherent networks, where the classical interconnect is realized through a wireless network-on-chip (WiNoC)~\cite{yazdanpanah_jsa22}.

In a WiNoC, the medium access control (MAC) protocol is a key determinant of performance and energy efficiency. Traditional WiNoC implementations commonly adopt a circulating-token mechanism, wherein a control token is passed sequentially among wireless nodes~\cite{abadal_ieeecm18}. The node holding the token gains exclusive access to the shared wireless channel for data transmission. While this approach ensures collision-free operation and fairness, it becomes inefficient under unbalanced or sparse traffic conditions. In particular, when the token reaches a node with no pending transmissions, the system wastes precious clock cycles waiting for the token to circulate to the next active node, leading to increased latency and suboptimal bandwidth utilization.

To address these inefficiencies, this paper introduces a novel MAC protocol that exploits the deterministic nature of quantum circuit execution. Unlike classical programs, quantum circuits are composed of a fixed, sequential set of operations without conditional branches. Consequently, the communication demands associated with circuit execution are known a priori. The proposed protocol, named \emph{instruction-directed token MAC (ID-TMAC)}, embeds this knowledge into the execution flow, allowing the token to be passed only among QCs that are scheduled to communicate, as determined by their predefined transmission order. This mechanism effectively eliminates idle token transitions and dynamically aligns medium access with the actual communication requirements dictated by the quantum algorithm.

\begin{figure}
    \centering
    \includegraphics[width=0.9\columnwidth]{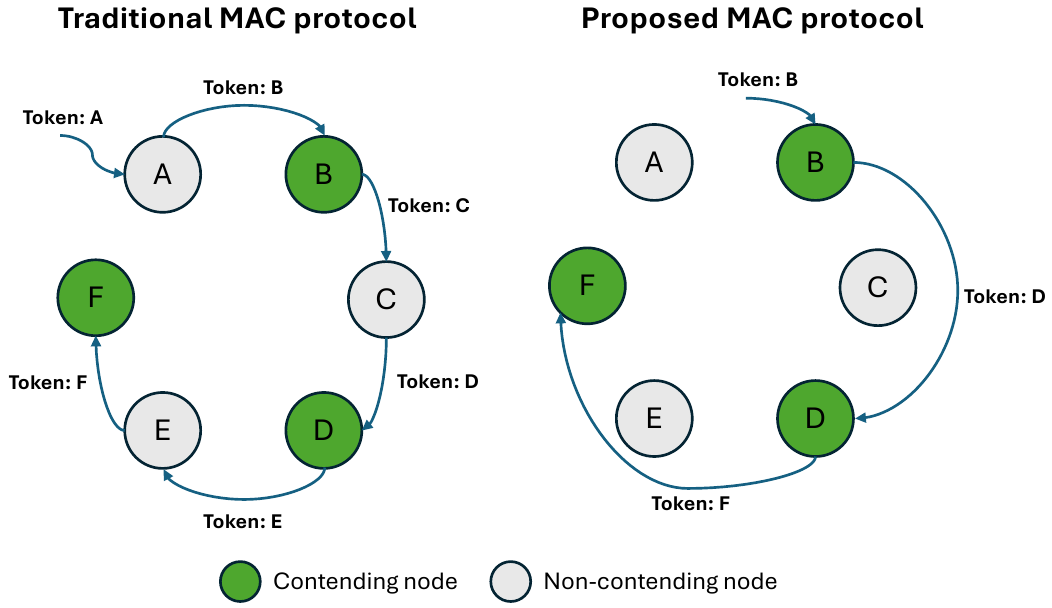}
    \caption{Traditional MAC protocol vs. proposed MAC protocol.}
    \label{fig:idea}
\end{figure}
As illustrated in Fig.~\ref{fig:idea}, the ID-TMAC approach contrasts with the baseline circulating token MAC (CT-MAC), where the token blindly traverses all nodes regardless of their communication status. By making MAC decisions instruction-aware and circuit-driven, ID-TMAC minimizes unnecessary token circulation, reduces communication latency, and improves overall radio-channel utilization, thereby enhancing the efficiency of the classical control plane in scalable multi-chip quantum architectures.


\section{Background on Quantum Teleportation}
In multi-core quantum architectures, computation frequently requires interactions between qubits that reside in different QCs. Implementing a two-qubit gate across physically separated qubits is non-trivial, as it would require either direct inter-core coupling (challenging to realize at scale) or the transfer of one qubit's state to the location of the other. The latter approach is particularly appealing, since it enables the execution of non-local gates while preserving the modularity of the architecture. However, physically moving a qubit between cores is impractical due to the fragility of quantum states and the constraints imposed by cryogenic integration.

A well-established method to overcome this limitation is the \emph{quantum teleportation protocol}~\cite{bose_prl03}. Teleportation allows the transfer of an unknown quantum state from a source qubit to a destination qubit without physically relocating the qubit itself. By leveraging entanglement and classical communication, teleportation enables the faithful transfer of quantum information across cores while respecting the architectural modularity of the system. As such, it provides a practical and scalable mechanism to implement inter-core quantum operations in distributed quantum processors.

\begin{figure}
    \centering
    \includegraphics[width=0.9\columnwidth]{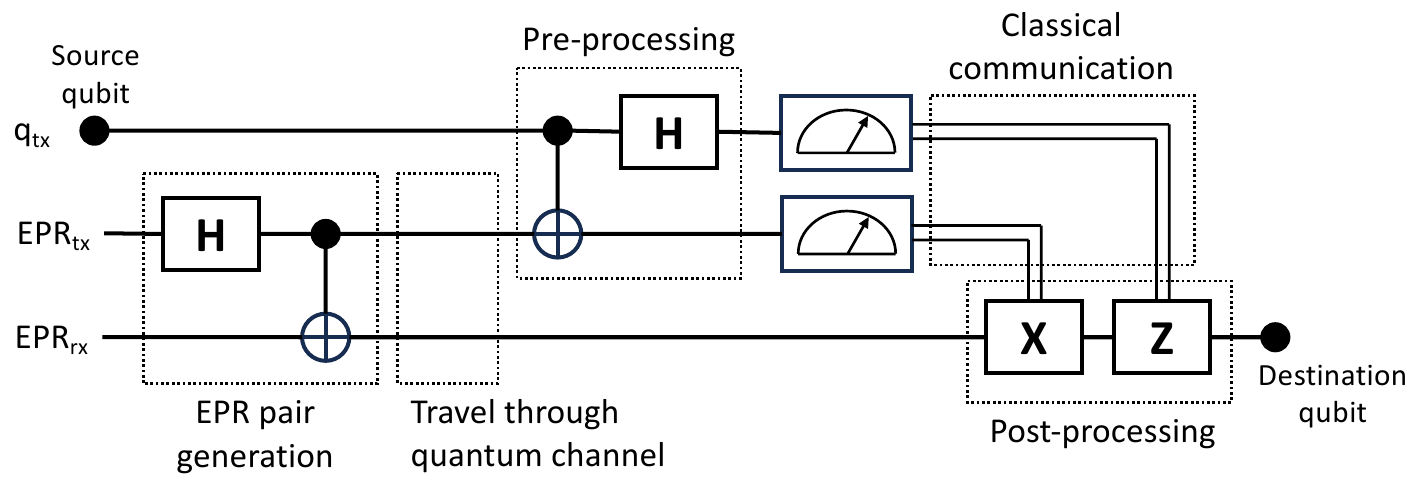}
    \caption{Steps involved in teleportation protocol.}
    \label{fig:teleportation}
\end{figure}
The teleportation protocol proceeds through five sequential phases: EPR generation, EPR distribution, pre-processing, classical communication, and post-processing, as illustrated in Fig.~\ref{fig:teleportation}. In the EPR generation phase, an entangled pair of qubits is created. These two entangled qubits, denoted as $\mathit{EPR}{tx}$ and $\mathit{EPR}{rx}$, are then distributed via quantum channels to the respective nodes where the source and destination qubits reside. During the pre-processing phase, the qubit to be transmitted ($q_{tx}$) and $\mathit{EPR}{tx}$ undergo a joint measurement, producing two classical correction bits. These bits are then transmitted to the destination node through a classical communication channel. Finally, in the post-processing phase, the received correction bits are used to determine which of four possible quantum operations should be applied to $\mathit{EPR}{rx}$, thereby reconstructing the state of $q_{tx}$ at the destination and completing the teleportation process.

\section{Reference Multi-Core Quantum System}

We consider a reference architecture representative of a fully cryogenically controlled multi-core quantum system, capturing the expected evolution of scalable quantum platforms where both qubits and control electronics operate at cryogenic temperatures to reduce latency and preserve coherence.

\begin{figure}
\centering
\includegraphics[width=0.99\columnwidth]{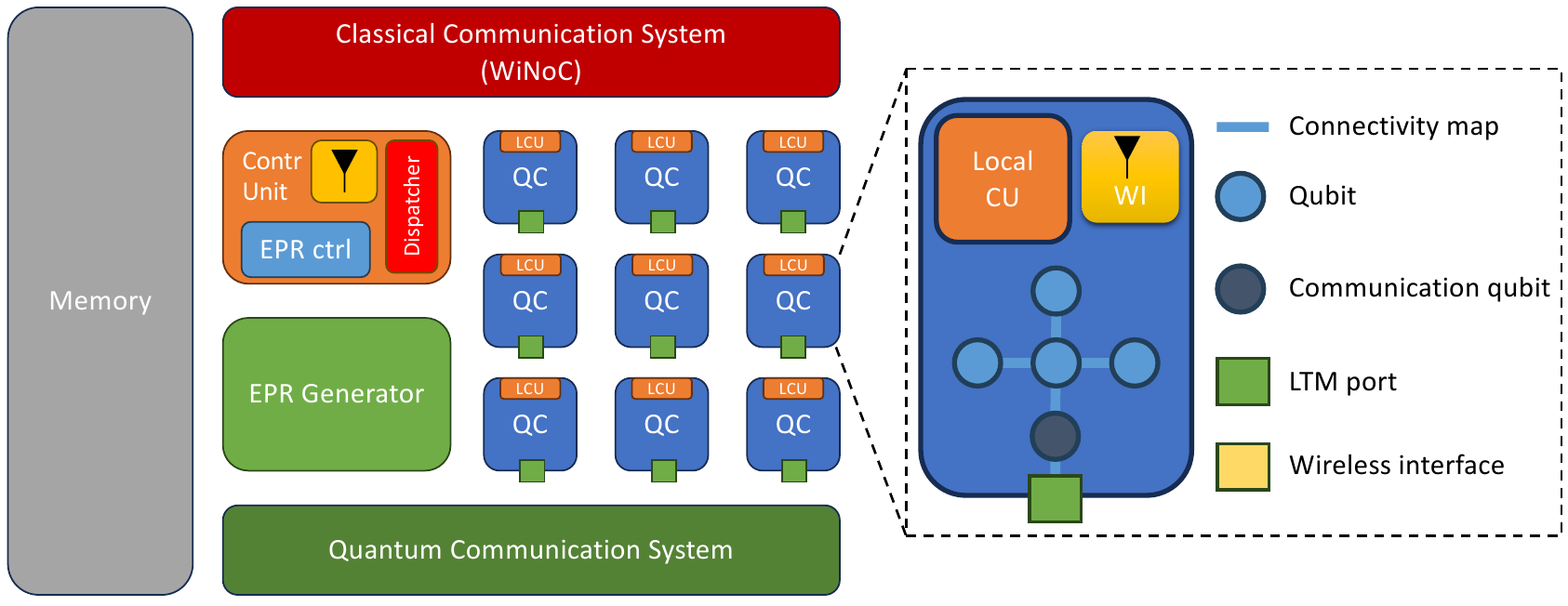}
\caption{Main modules of the reference multi-core quantum architecture.}
\label{fig:modules_and_qcore}
\end{figure}

As shown in Fig.~\ref{fig:modules_and_qcore}, the architecture comprises the following key modules: (i) a \emph{Memory} that stores program instructions, (ii) a global \emph{Control Unit} (CU) that fetches, decodes, and dispatches instruction bundles, (iii) an \emph{EPR Generator} for creating and distributing entangled pairs to support teleportation, (iv) an array of \emph{Quantum Cores} that execute quantum gates, and (v) a \emph{Communication System} composed of a quantum and a classical plane. The quantum plane transfers qubit states, while the classical plane coordinates control, synchronization, and teleportation post-processing.

Each QC integrates a set of physical qubits, one or more light-to-matter (LTM) ports, and a local controller. LTM ports enable interaction with remote QCs through photonic channels by using dedicated \emph{communication qubits}. The local controller interprets instructions received from the global CU and drives local operations according to the QC's connectivity map.

\begin{figure*}
    \centering
    \includegraphics[width=0.90\textwidth]{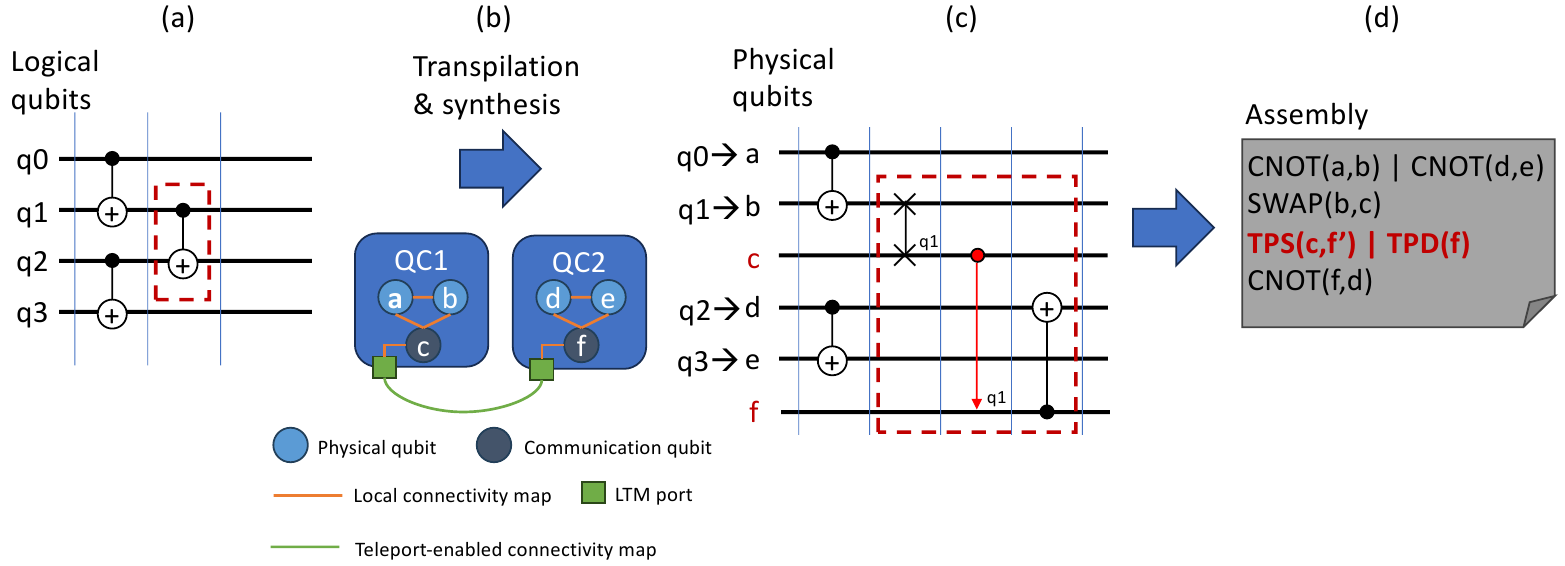}
    \caption{From circuit to assembly code. Logical circuit (a). Compilation phase (b). Synthesized circuit (c). Assembly code (d).}
    \label{fig:circuit}
\end{figure*}
The compilation flow maps the logical quantum circuit to physical qubits while respecting both local and teleport-enabled connectivity (Fig.~\ref{fig:circuit}). Operations that involve qubits belonging to different QCs are transformed into teleportation-based instructions. The resulting assembly code consists of \emph{instruction bundles}, each grouping operations that can be executed in parallel. Two special instructions are introduced: \texttt{TPS} for the source QC, handling the generation, distribution, and measurement phases of teleportation; and \texttt{TPD} for the destination QC, which performs the final correction based on the received classical bits.

\begin{figure}
\centering
\includegraphics[width=0.90\columnwidth]{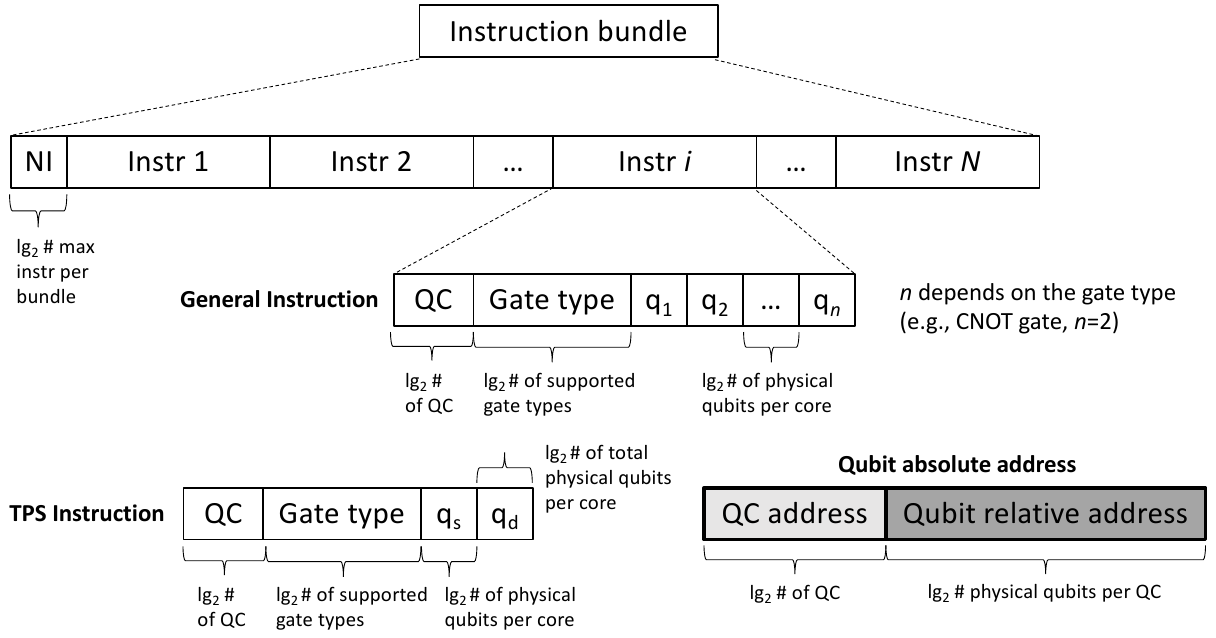}
\caption{Instruction bundle structure and qubit addressing.}
\label{fig:bundle}
\end{figure}
An instruction bundle, Fig.~\ref{fig:bundle}, encodes the number of instructions followed by a list of operations. Each instruction specifies the QC address, gate type, and operands' local addresses. The CU fetches bundles from memory, decomposes them into individual instructions, and dispatches them to the target QCs. When teleportation is required, the CU configures the EPR generator accordingly and coordinates the exchange of the classical correction bits through the classical interconnect.

\begin{figure}
    \centering
    \includegraphics[width=0.8\columnwidth]{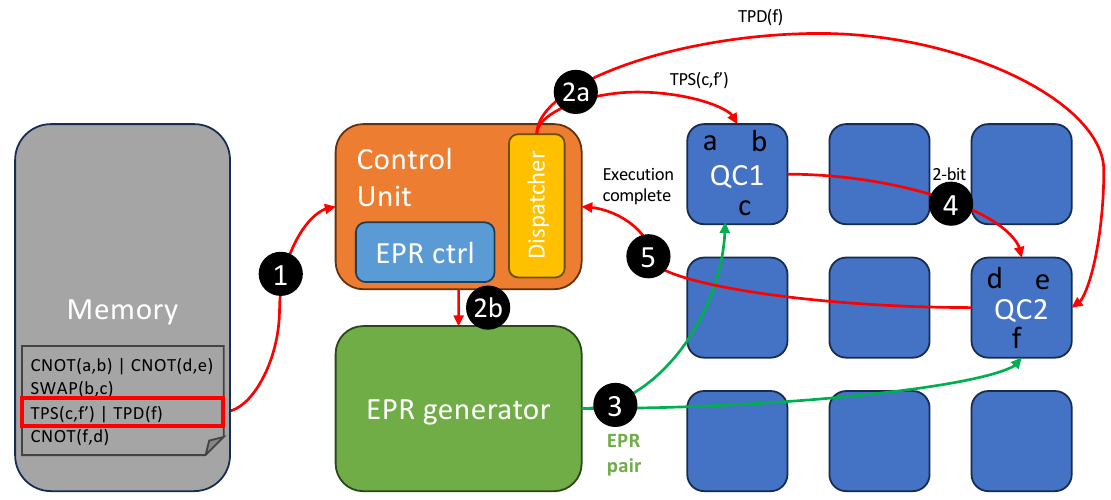}
    \caption{Phases involved in the execution of the remote bundle $\langle$\texttt{TPS(c,f) | TPD(f)}$\rangle$.}
    \label{fig:tpstpd}
\end{figure}
Classical communication among the CU and QCs is realized via a WiNoC, which ensures scalable and low-latency control at cryogenic temperatures~\cite{sebastiano_sscm25}. The EPR generator interfaces with the QCs through dedicated quantum channels connected to their LTM ports. 


The execution model is summarized in Fig.~\ref{fig:tpstpd} for the red coloured bundle. The CU fetches the instruction bundle $\langle$\texttt{TPS(c,f') | TPD(f)}$\rangle$, which implements the teleportation operation, aiming to teleport the quantum state of $c$ to $f$ (Fig.~\ref{fig:tpstpd}\cone). The \texttt{TPS} instruction is dispatched to QC1, and the \texttt{TPD} instruction is sent to QC2 (Fig.~\ref{fig:tpstpd}\ctwo a). This time, the address of the second operand in \texttt{TPS} is not replaced with its local address, as QC1 needs to know the address of $f$ in order to send the two classical bits of information from QC1 to QC2 during the first part of the teleportation protocol. Simultaneously, the EPR control module configures the EPR generator (Fig.~\ref{fig:tpstpd}\ctwo b), which produces the EPR pair distributed to both QC1 and QC2 (Fig.~\ref{fig:tpstpd}\cthree). The local CU of QC1 executes the pre-processing phase of the teleportation protocol and transmits the 2-bit classical information to QC2 (Fig.~\ref{fig:tpstpd}\cfour). The CU of QC2 then performs the post-processing phase to complete the teleportation. Finally, it sends an \emph{execution complete} message to the CU (Fig.~\ref{fig:tpstpd}\cfive), allowing it to fetch the next instruction.


\section{MAC Protocol}

\subsection{Traditional MAC Protocol}
In wireless communication systems, the MAC layer regulates access to the shared radio medium to prevent collisions and ensure efficient bandwidth utilization. In the context of WiNoC architectures, where on-chip wireless interfaces (WIs) enable long-range, low-latency communication among processing cores, the MAC protocol plays a central role in coordinating transmissions across multiple wireless nodes.

The traditional MAC scheme adopted in WiNoCs is based on token-controlled medium access, commonly referred to as the Circulating Token MAC (CT-MAC). In this approach, a unique control packet (called a token) is circulated among all wireless nodes connected to the shared channel. At any given time, only the node holding the token is permitted to transmit data. After completing its transmission, or once a predefined time slot expires, the token is released and forwarded to the next node in a predetermined sequence. This mechanism guarantees collision-free communication and simplifies arbitration, as only one node can access the channel at a time.

The CT-MAC protocol, while conceptually simple, presents several drawbacks when applied to WiNoC-based communication systems. Its performance is highly dependent on token circulation latency and the spatial distribution of traffic. Under bursty or uneven traffic conditions, the fixed token-passing order can cause significant delays, as active nodes must wait for the token to traverse multiple idle nodes before gaining access to the medium. This leads to inefficient utilization of the wireless channel and increased communication latency. As a result, CT-MAC becomes progressively less efficient as system size and communication irregularity increase, highlighting the need for more adaptive and traffic-aware arbitration mechanisms.

\subsection{Proposed MAC Protocol}
The proposed MAC protocol builds on the observation that, after compilation, the complete communication schedule of a quantum circuit is known in advance. In particular, the compiler identifies all teleportation operations and determines the corresponding transmission events between QCs. This knowledge allows embedding, directly within the instructions, information that specifies whether a QC must access the wireless medium during the execution of that instruction and, if so, in which order relative to other participating QCs.

Leveraging this insight, the proposed protocol---referred to as the Instruction-Directed Token MAC (ID-MAC)---optimizes channel arbitration by circulating the token only among QCs that are scheduled to transmit. Consequently, token handovers to idle or non-contending nodes are avoided, reducing unnecessary token circulation and improving overall communication efficiency.

\subsubsection{Packet Formats}
\begin{figure}
    \centering
    \includegraphics[width=0.9\columnwidth]{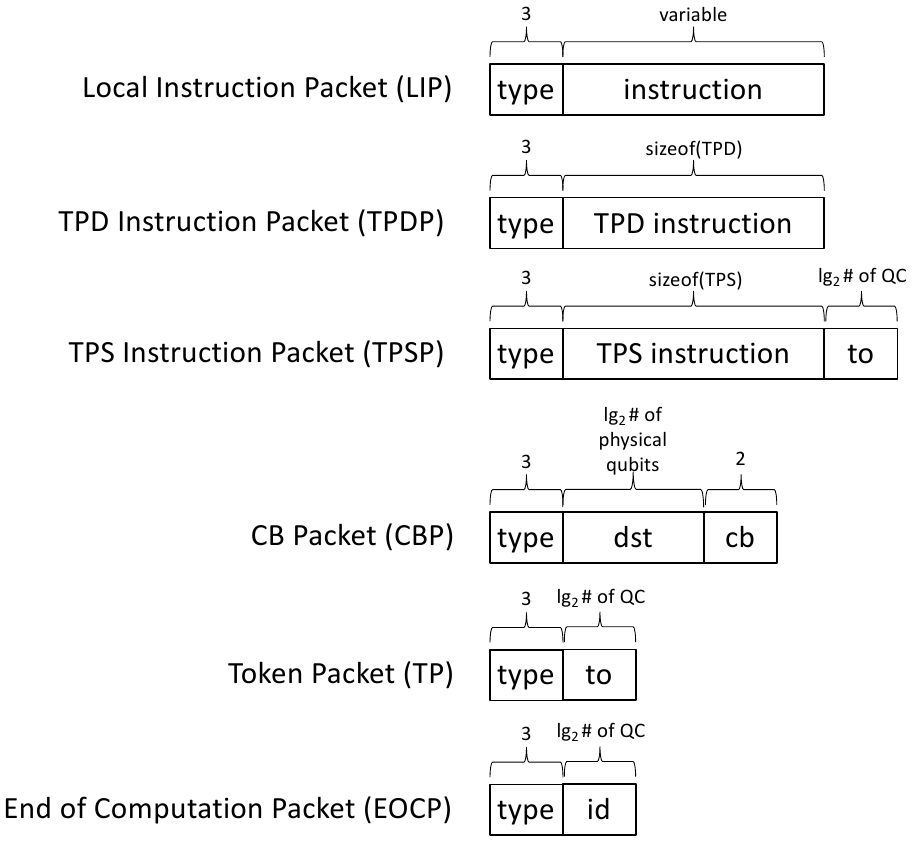}
    \caption{Packet formats supporting the ID-MAC protocol.}
    \label{fig:packet_format}
\end{figure}
To support ID-MAC operation, several packet formats are defined, as illustrated in Fig.~\ref{fig:packet_format}. The first three bits encode the packet type, while the remaining fields depend on the packet's purpose. The Local Instruction Packet (LIP) is used by the dispatcher to distribute general, non-teleportation instructions to the QCs. The TPS Instruction Packet (TPSIP) carries teleportation-source (\texttt{TPS}) instructions and includes an additional \emph{token order} (\emph{to}) field, which specifies the sequence in which the corresponding teleportation operation is permitted to access the wireless medium. The TPD Instruction Packet (TPDIP) contains teleportation-destination (\texttt{TPD}) instructions, which are stored locally at the target QC. The Correction Bits Packet (CBP) conveys the classical correction bits (\emph{cb}) generated during the pre-processing phase of the teleportation protocol from the source QC to the destination QC. Finally, the Token Packet (TP) represents the circulating token itself and includes the \emph{to} field that identifies the next QC authorized to use the radio medium.

\subsubsection{Protocol Operation}
The ID-MAC operation unfolds across two phases: the dispatch phase and the execution phase.

\paragraph{Dispatch phase}
During this phase, the Dispatch process (Alg.~\ref{alg:dispatch}) is executed by the global Dispatch Unit. It receives an instruction bundle as input, generates the corresponding instruction packets (LIP, TPSIP, or TPDIP) using the MakeInstructionPacket function, and transmits them sequentially.
For each \texttt{TPS} instruction, the function assigns a token order value (\emph{to}), starting from zero and incrementing for subsequent instructions. Since only the dispatcher transmits during this phase, no medium contention occurs.
\begin{algorithm}
\small
\caption{Dispatch process executed by the Dispatch unit during the dispatch phase}
\label{alg:dispatch}
\begin{algorithmic}[1]
\Process{Dispatch}{Bundle $b$}
    \State \emph{instr} $\gets$ \Call{GetFirstInstruction}{$b$}
    \Repeat
        \State \emph{pkt} $\gets$ \Call{MakeInstructionPacket}{\emph{instr}}
        \State \Call{Transmit}{\emph{pkt}}
        \State \emph{instr} $\gets$ \Call{GetNextInstruction}{$b$}
    \Until{\emph{instr} = NULL}
\EndProcess
\end{algorithmic}
\end{algorithm}

Each QC executes the BufferFillUp process (Alg.~\ref{alg:buffer_fillup}), which receives instruction packets and stores them in the appropriate local buffers: local instructions in \emph{LIB}, teleportation-destination instructions in \emph{TPDB}, and teleportation-source instructions in \emph{TPSB}. The TPSB buffer is sorted according to the token order field.
\begin{algorithm}
\small
\caption{BufferFillUp process executed by the LCU of each QC during the dispatch phase}
\label{alg:buffer_fillup}
\begin{algorithmic}[1]
\Process{BufferFillUp}{}
    \State \Call{Receive}{\emph{pkt}}
    \If{\emph{pkt.type} = LIP}
        \State \Call{Push}{LIB, \emph{pkt.instr}}
    \ElsIf{\emph{pkt.type} = TPDIP}
        \State \Call{Push}{TPDB, \emph{pkt.instr}}
    \ElsIf{\emph{pkt.type} = TPSIP}
        \State \Call{InsertSort}{TPSB, \emph{pkt.instr}, \emph{pkt.to}}
    \EndIf
\EndProcess
\end{algorithmic}
\end{algorithm}

\paragraph{Execution Phase}
During execution, each QC runs four concurrent processes: LocalInstruction, TPSInstruction, TPDInstruction, and EndOfComputation.

The LocalInstruction process (Alg.~\ref{alg:local_instruction}) repeatedly fetches and executes instructions from the LIB until the buffer is empty.
\begin{algorithm}
\small
\caption{LocalInstruction process executed by the LCU in QC unit during the execution phase}
\label{alg:local_instruction}
\begin{algorithmic}[1]
\Process{LocalInstruction}{LIB}
    \While{LIB is not empty}
        \State \emph{instr} $\gets$ \Call{Pop}{LIB}
        \State \Call{Execute}{\emph{instr}}
    \EndWhile
\EndProcess
\end{algorithmic}
\end{algorithm}

The TPSInstruction process (Alg.~\ref{alg:tps_instruction}) manages teleportation-source operations that require access to the shared radio channel. For each instruction in TPSB, the process waits for the token corresponding to its \emph{to} value (skipping this step if \emph{to} = 0). After executing the instruction, the generated correction bits are encapsulated in a CBP and transmitted. The QC then increments the token order and transmits a new token to grant access to the next scheduled QC.
\begin{algorithm}
\small
\caption{TPSInstruction process managing teleportation-source operations. The process is executed by the LCU in QC unit during the execution phase}
\label{alg:tps_instruction}
\begin{algorithmic}[1]
\Process{TPSInstruction}{TPSB}
    \While{TPSB is not empty}
        \State \emph{instr}, \emph{to} $\gets$ \Call{Pop}{TPSB}
        \If{\emph{to} $\ne$ 0}
            \State \Call{ReceiveToken}{\emph{to}}
        \EndIf
        \State \emph{cb} $\gets$ \Call{Execute}{\emph{instr}}
        \State \emph{pkt} $\gets$ \Call{MakeCBPacket}{\emph{cb}, \emph{instr}}
        \State \Call{Transmit}{\emph{pkt}}
        \State \emph{pkt} $\gets$ \Call{MakeTokenPacket}{\emph{to}+1}
        \State \Call{Transmit}{\emph{pkt}}
    \EndWhile
\EndProcess
\end{algorithmic}
\end{algorithm}

The TPDInstruction process (Alg.~\ref{alg:tpr_instruction}) executes the post-processing phase of teleportation. It waits for the arrival of the correction bits, then applies the corresponding quantum operation to complete the teleportation.
\begin{algorithm}
\small
\caption{TPDInstruction process handling teleportation-destination operations. The process is executed by the LCU in QC unit during the execution phase}
\label{alg:tpr_instruction}
\begin{algorithmic}[1]
\Process{TPDInstruction}{TPRB}
    \pFor{\emph{instr} $\in$ TPRB}
        \State \emph{cb} $\gets$ \Call{ReceiveCorrectionBits}{ }
        \State \Call{Execute}{\emph{instr}, \emph{cb}}
    \EndFor
\EndProcess
\end{algorithmic}
\end{algorithm}

Finally, the EndOfComputation process (Alg.~\ref{alg:eoc}) monitors the completion of all local operations. Once the LIB, TPDB, and TPSB buffers are empty, it sends an end-of-computation packet to the Control Unit.
\begin{algorithm}
\small
\caption{EndOfComputation process executed by the LCU in QC unit during the execution phase}
\label{alg:eoc}
\begin{algorithmic}[1]
\Process{EndOfComputation}{LIB, TPDB, TPSB}
    \If{LIB and TPDB and TPSB are empty}
        \State \emph{pkt} $\gets$ \Call{MakeEOCPacket}{self.\emph{qcore\_id}}
         \State \Call{Transmit}{\emph{pkt}}       
    \EndIf
\EndProcess
\end{algorithmic}
\end{algorithm}

\section{Experiments}
In this section, we present a set of experiments designed to (i) motivate the importance of the classical communication subsystem in scalable multi-core quantum architectures and (ii) evaluate the performance of the proposed ID-MAC protocol. All experiments are conducted using the \qcomm{} framework~\cite{palesi_qcomm25}. The main simulation parameters and configuration details are summarized in Tab.~\ref{tab:parameters}.
\begin{table}
    \centering
    \caption{Physical parameters, architectural and micro-architectural parameters, and circuit properties used for the experiments.}
    \label{tab:parameters}
    \begin{tabular}{lc}
        \toprule
        \multicolumn{2}{c}{Physical parameters and microarchitectural parameters} \\
        \midrule
        Mean of EPR pair generation time    & $10^3$ \si{\nano\second}$^{*}$ \\
        EPR pair distribution time          & 0.01 \si{\nano\second}$^{*}$ \\
        Pre-processing time                 & 390 \si{\nano\second}$^{*}$ \\
        Post-processing time                & 30 \si{\nano\second}$^{*}$ \\
        WiNoC bitrate                       & 12 Gbps \\
        Token pass time                     & 1 \si{\nano\second} \\
        RAM bandwidth                       & 128 Gbps$^{**}$ \\ 
        Bits per instruction                & 4$^{\dag}$\\
        Decode time                         & $10$ \si{\nano\second} per instruction\\
        \midrule
        \multicolumn{2}{c}{Architectural parameters and circuit properties} \\
        \midrule
        Network topology & mesh \\
        Mesh size & from $1 \times 2$ to $10 \times 10$ \\
        Number of QCs & from 1 to 100 \\
        LTM ports per QC & 1 \\
        Physical qubits & from 10 to 10,000 \\
        Circuit size -- logical qubits & from 10 to 1000\\
        Circuit size -- gates & from 100 to 10,000\\
        \bottomrule
        \multicolumn{2}{p{0.8\linewidth}}{$^{*}$Data from~\cite{morten_arcmp20}. $^{**}$DDR4 SDRAM. $^{\dag}$The instruction set is assumed to consist of up to 16 instructions.}
    \end{tabular}
\end{table}

\subsection{Assessing the Role of Classical Communication}
We first evaluate the impact of the classical communication subsystem on the overall execution time of the multi-core quantum architecture. To this end, we measure the fraction of total execution time spent on classical communication activities. The analysis is carried out using randomly generated quantum circuits whose size (i.e., number of qubits and gates) scales with the number of quantum cores (QCs).

We consider homogeneous systems, where each QC hosts 16 physical qubits. For a system comprising $n$ QCs, the corresponding random circuit contains $16 \times n$ qubits and $160 \times n$ gates, evenly divided between single-qubit and two-qubit operations. No optimized qubit mapping is applied; instead, logical qubit $i$ is assigned to QC $i \bmod n$, ensuring a uniform distribution of operations across cores.

\begin{figure}
    \centering
    \includegraphics[width=0.9\columnwidth]{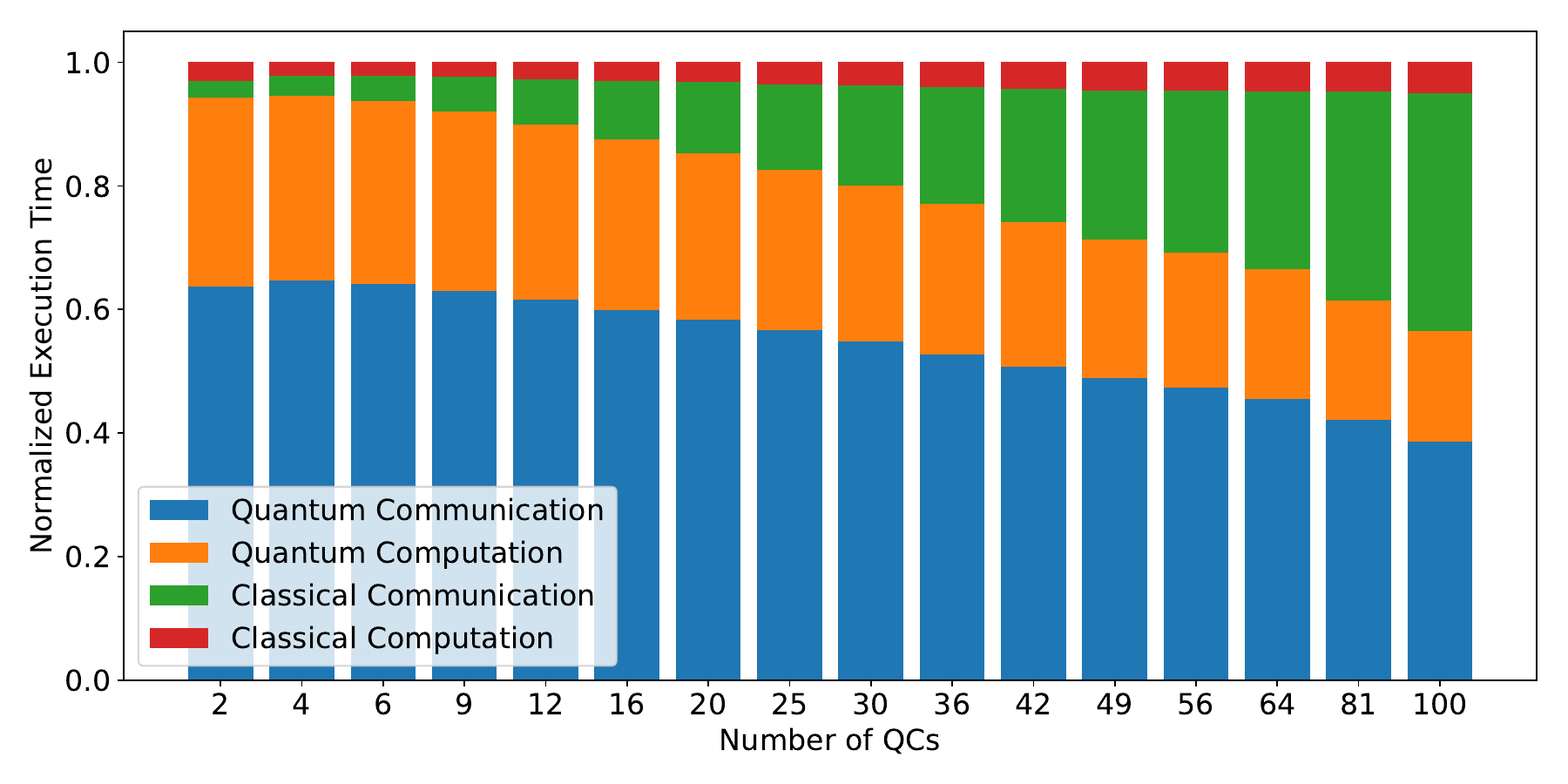}
    \caption{Breakdown of the normalized execution time as a function of system size.}
    \label{fig:etime_vs_size}
\end{figure}
Fig.~\ref{fig:etime_vs_size} reports the normalized execution time breakdown as the number of QCs increases from 1 to 100. The total execution time is decomposed into four components: \emph{Quantum communication}, time spent on quantum-state transfer operations, including EPR generation and distribution;
\emph{Quantum computation}, time required for gate execution and the pre-/post-processing phases of the teleportation protocol;
\emph{Classical communication}, time spent transmitting control and data packets, such as instruction bundles, teleportation correction bits, and end-of-computation notifications; and
\emph{Classical computation}, time consumed by local control tasks, including instruction decoding and scheduling. As observed in Fig.~\ref{fig:etime_vs_size}, the contribution of classical communication (green segment) becomes increasingly significant as the number of QCs grows. This trend reflects the expanding need for coordination and data exchange across cores as the system scales.

\begin{figure}
    \centering
    \includegraphics[width=0.9\columnwidth]{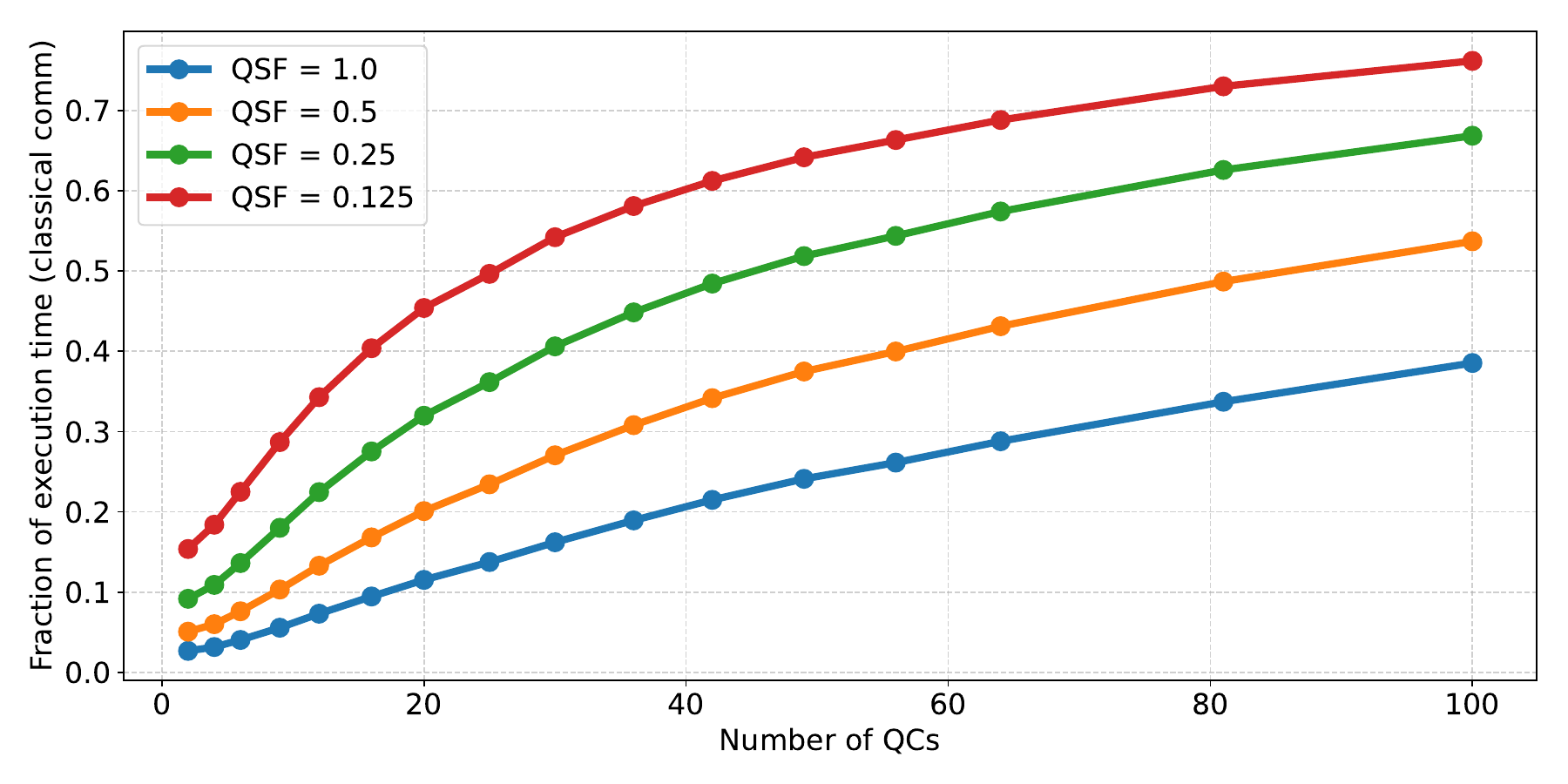}
    \caption{Fraction of execution time spent in classical communication for different quantum scaling factors (QSFs) and system sizes.}
    \label{fig:etime_vs_size_vs_qsf}
\end{figure}
To further explore this trend, we analyze how improvements in quantum technology affect the relative importance of classical communication. We define a \emph{Quantum Scaling Factor} (QSF), which scales all quantum-related delays (e.g., gate latency, EPR generation time) by a constant factor to emulate technology advancement. Fig.~\ref{fig:etime_vs_size_vs_qsf} shows the fraction of execution time devoted to classical communication for varying numbers of QCs and QSF values. As the QSF decreases—representing faster quantum operations—the fraction of time spent in classical communication increases.

This behavior highlights a fundamental architectural insight: as quantum technologies evolve and quantum operations become faster, the latency gap between quantum and classical components narrows, eventually making classical communication a dominant performance bottleneck. Furthermore, this effect is exacerbated as the system scales to a larger number of QCs. These results emphasize the necessity of optimizing the classical communication subsystem. In the following subsection, we demonstrate how the proposed ID-MAC effectively alleviates this emerging bottleneck.

\subsection{Assessment of ID-MAC}
We now evaluate the effectiveness of the proposed ID-MAC protocol by comparing it against the baseline CT-MAC. The goal is to quantify the reduction in classical communication overhead and its impact on total execution time across different system scales and quantum technology assumptions. The analysis considers varying numbers of QCs and QSFs. For each configuration, we measure: (i) the fraction of total execution time devoted to classical communication when using CT-MAC and ID-MAC, and (ii) the overall reduction in execution time achieved by ID-MAC.

\begin{figure*}
\centering
    \begin{subfigure}{0.45\textwidth}
    \centering
    \includegraphics[width=\linewidth]{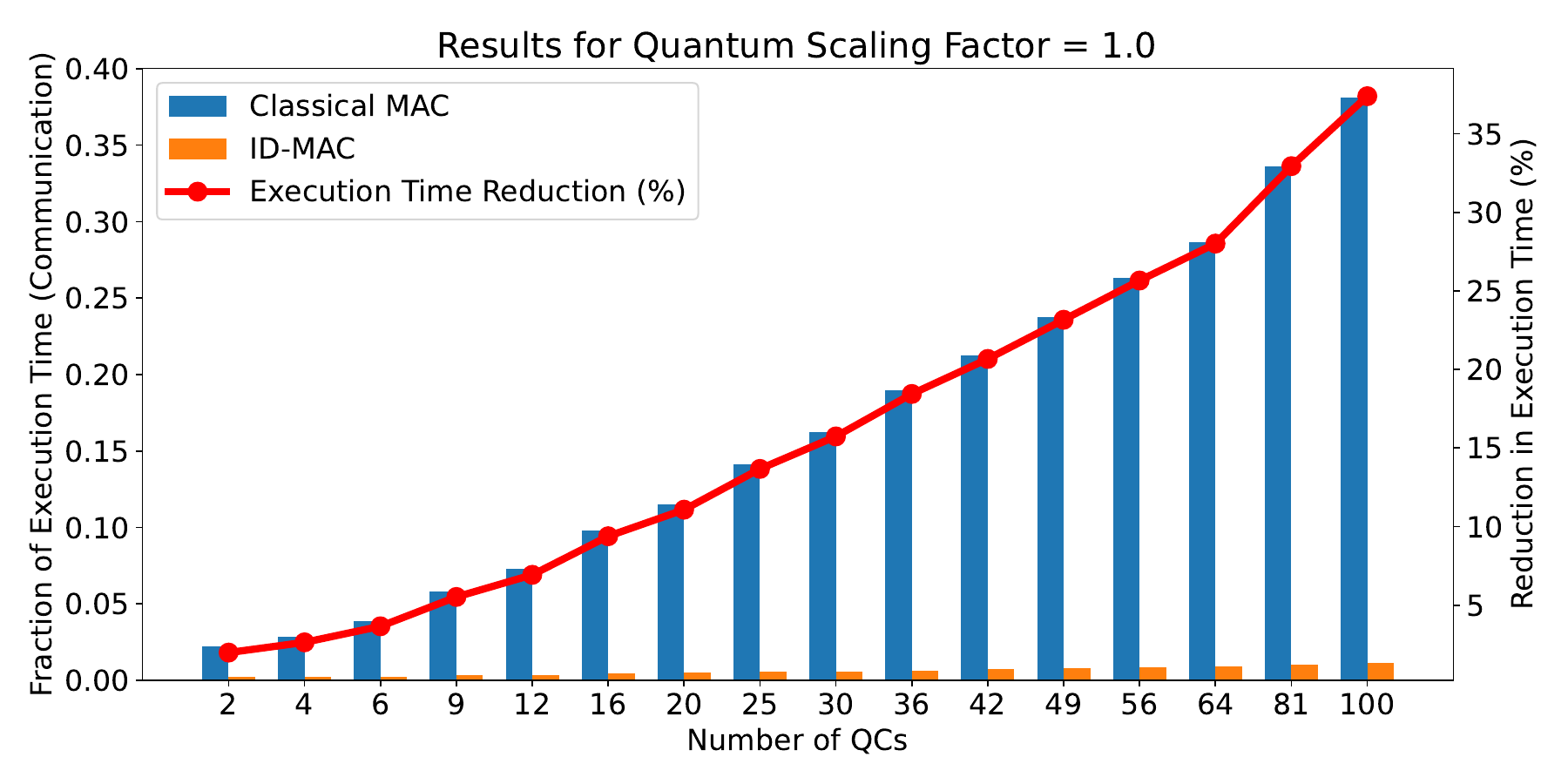}
    \end{subfigure} \hfill
    \begin{subfigure}{0.45\textwidth}
    \centering
    \includegraphics[width=\linewidth]{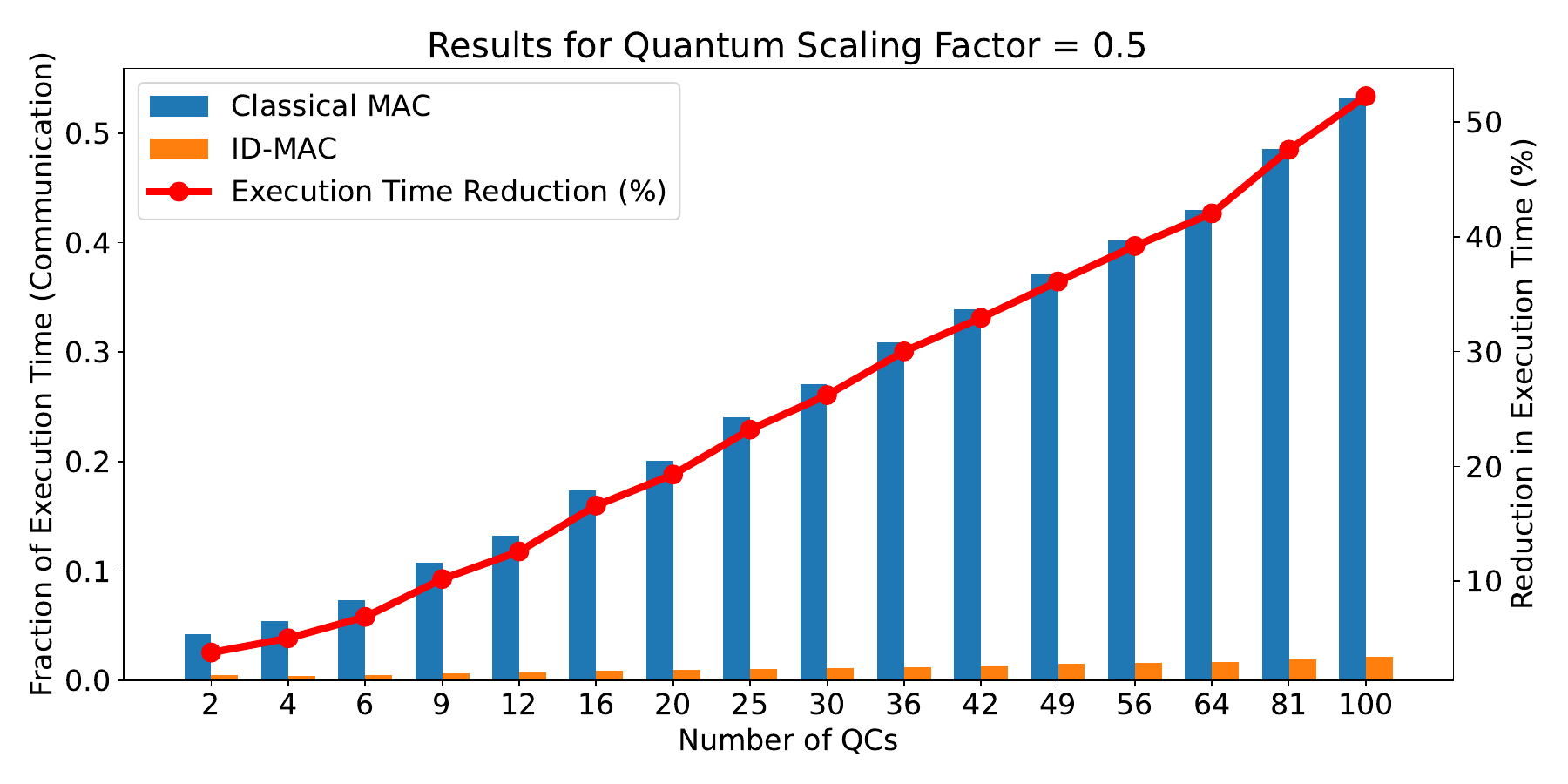}
    \end{subfigure}


    \begin{subfigure}{0.45\textwidth}
    \centering
    \includegraphics[width=\linewidth]{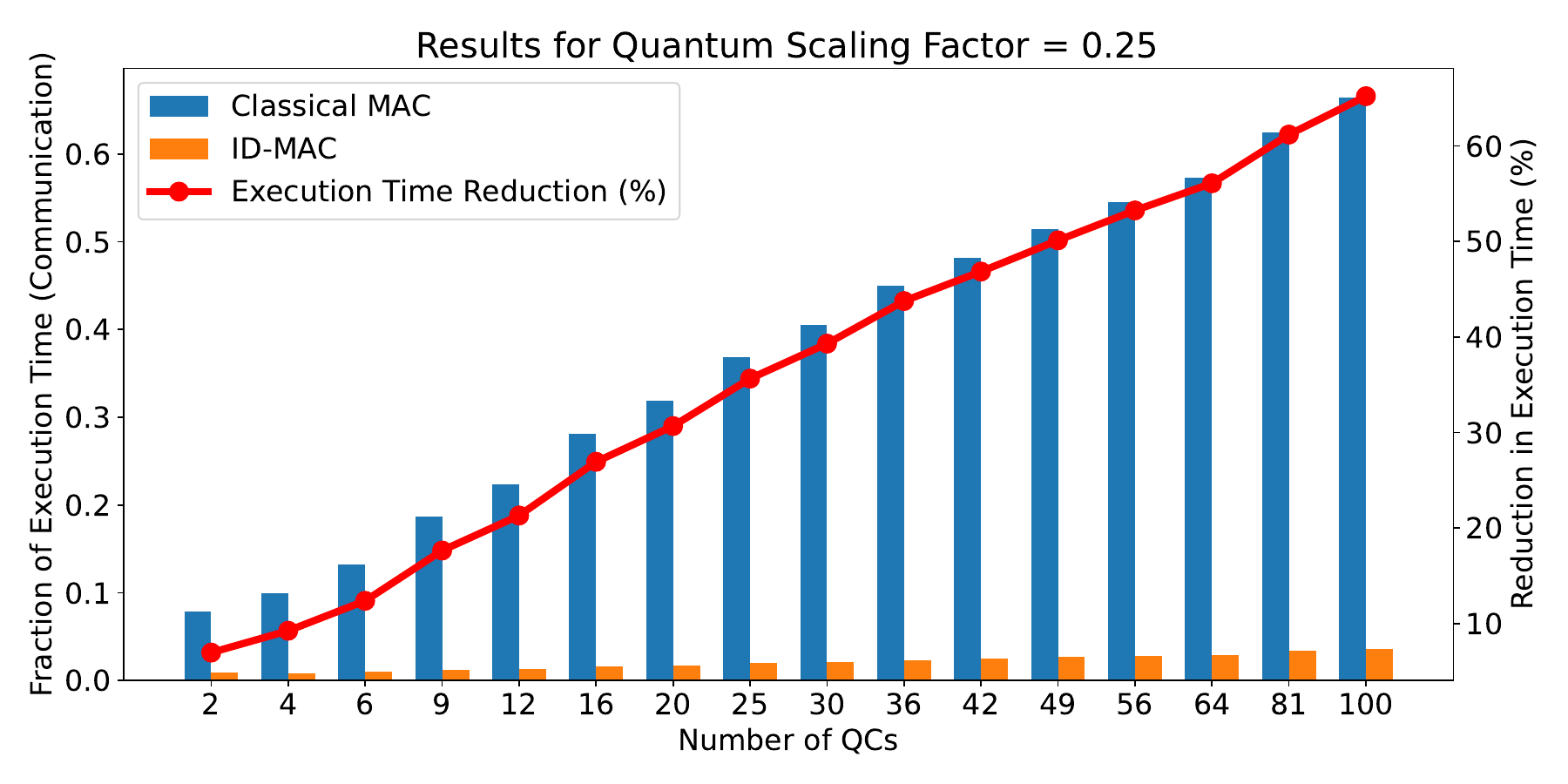}
    \end{subfigure}\hfill
    \begin{subfigure}{0.45\textwidth}
    \centering
    \includegraphics[width=\linewidth]{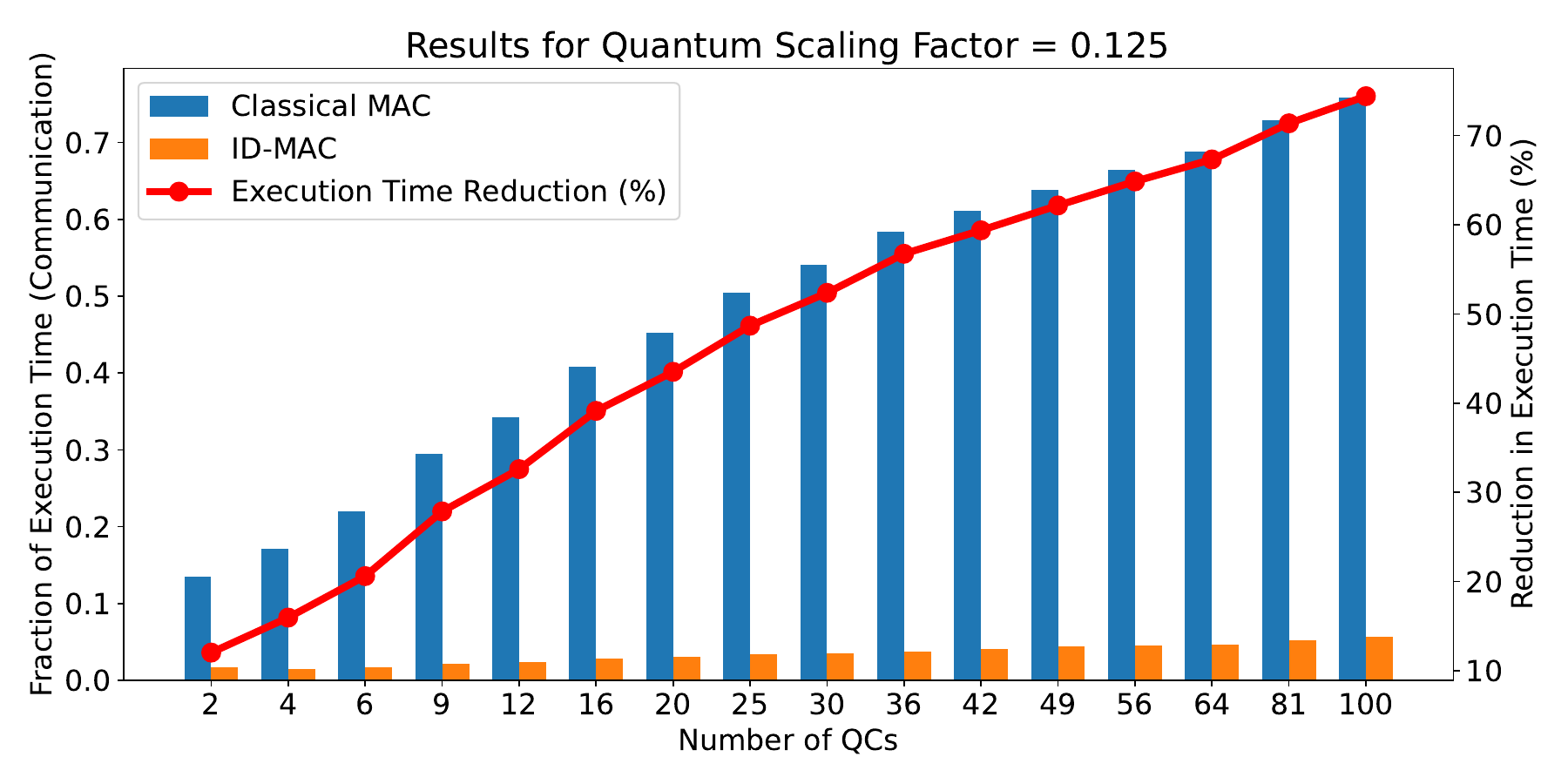}
    \end{subfigure}
\caption{Fraction of execution time spent in classical communication and overall execution time savings obtained using ID-MAC for different system sizes and QSFs.}
\label{fig:etime_saving}
\end{figure*}
Fig.~\ref{fig:etime_saving} summarizes the results. ID-MAC consistently achieves a substantial reduction in the time spent on classical communication compared to CT-MAC. The improvement becomes more pronounced as the system scales, since the traditional token-based protocol suffers increasingly from idle token transitions among non-transmitting QCs. By contrast, ID-MAC restricts token circulation to active transmitters only, significantly improving medium utilization efficiency. This reduction in communication latency directly translates into lower overall execution times. The red curves in Fig.~\ref{fig:etime_saving} report the total execution time savings when switching from CT-MAC to ID-MAC. For a system with 100 QCs, the proposed protocol achieves speedups of approximately 35\%, 50\%, 65\%, and 75\% for QSF values of 1.0, 0.5, 0.25, and 0.125, respectively. These results confirm that the benefits of ID-MAC amplify as both system size increases and quantum operations become faster—conditions representative of future large-scale quantum systems.

To further validate the results, we also assess ID-MAC using a diverse set of real-world quantum benchmark circuits from the Qiskit framework~\cite{qiskit2024} and MQTBench~\cite{quetschlich2023mqtbench}. The selected benchmarks include amplitude estimation (ae), Greenberger–Horne–Zeilinger (ghz) state preparation, graph state (graphstate) generation, quantum Fourier transform (qft), quantum neural network (qnn), and a randomly generated circuit. Each circuit is transpiled and optimized using Qiskit to target a native gate set composed of Z-rotations, $\sqrt{X}$, X, and controlled-X operations. All benchmarks, generated with 25 qubits, are executed on a system comprising four QCs, each containing nine physical qubits. Qubit mapping is performed using TeleSABRE~\cite{russo_qce25}, an extension of the SABRE heuristic~\cite{li_asplos19} specifically tailored for multi-core quantum processors. 

\begin{figure}
    \centering
    \includegraphics[width=0.9\columnwidth]{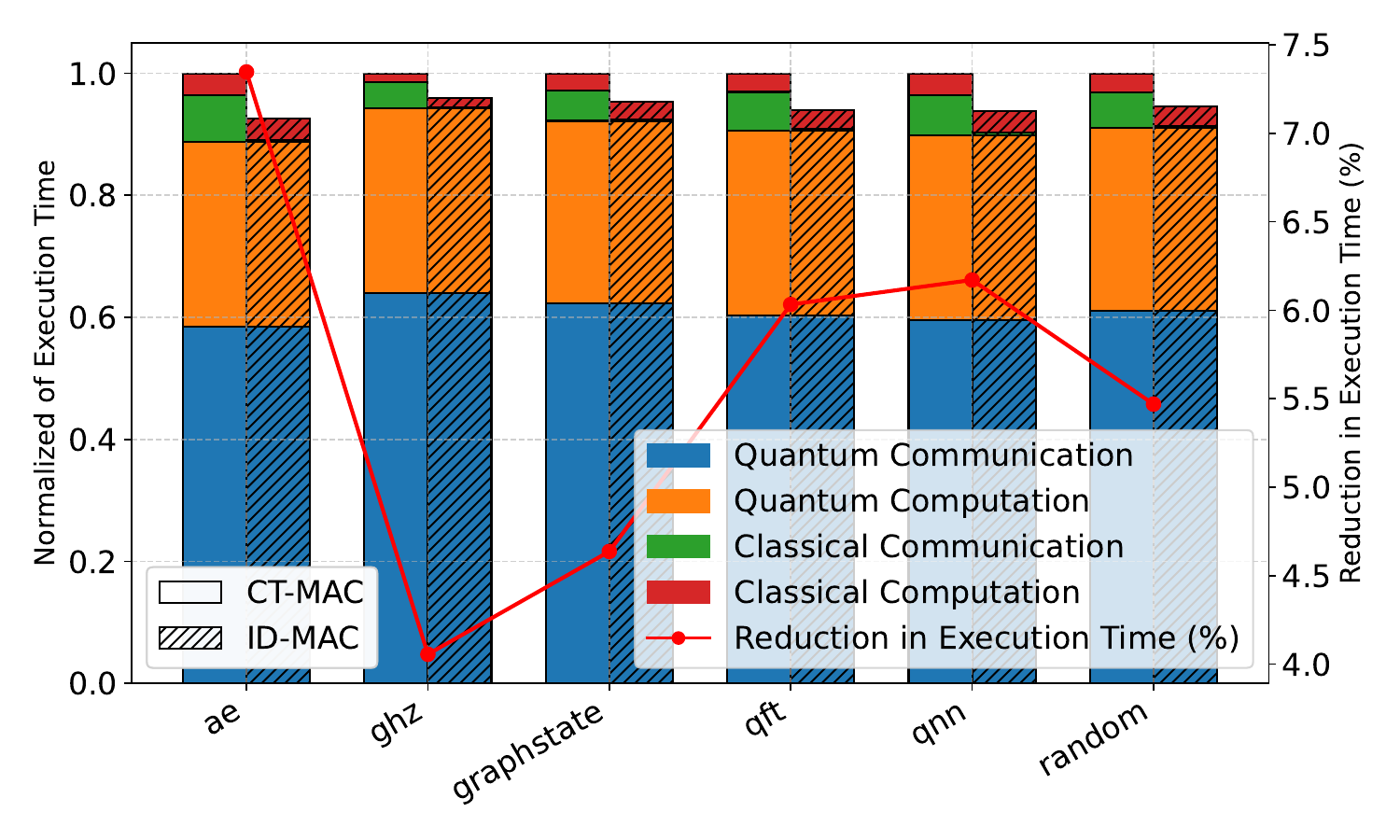}
    \caption{Comparison between CT-MAC and ID-MAC for real benchmark circuits: total execution time and percentage improvement.}
    \label{fig:real_benchmarks}
\end{figure}
Fig.~\ref{fig:real_benchmarks} reports the comparison between CT-MAC and ID-MAC for these benchmarks. Although the overall system size (four QCs) limits the absolute contribution of classical communication to total execution time, ID-MAC still provides measurable improvements. Specifically, the proposed protocol nearly eliminates classical communication delays, yielding total execution time reductions between 4\% and 7\%, depending on the circuit.

\begin{figure}
    \centering
    \includegraphics[width=0.9\columnwidth]{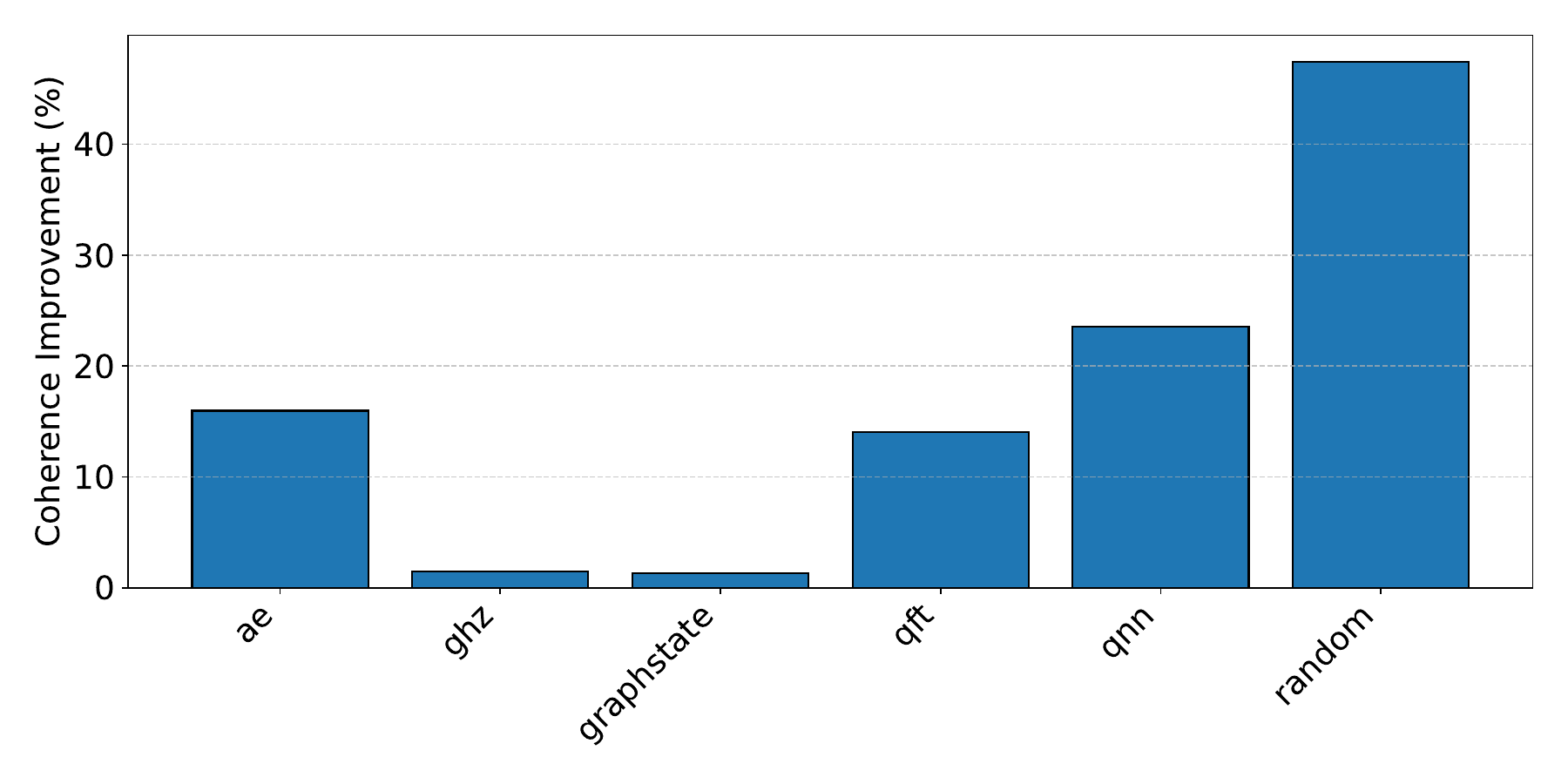}
    \caption{Improvement in effective coherence time when ID-MAC is adopted compared to CT-MAC.}
    \label{fig:coherence_improvement}
\end{figure}
Fig.~\ref{fig:coherence_improvement} illustrates the improvement in effective coherence time obtained when adopting ID-MAC instead of CT-MAC. The coherence time is estimated using the analytical model proposed in~\cite{escofet_qce25}, which relates the overall system coherence to the execution latency of quantum operations. Since ID-MAC reduces the total execution time by minimizing classical communication overheads, it effectively prolongs the usable coherence window of the system. As shown in the figure, even moderate reductions in execution time translate into substantial gains in effective coherence. 

These findings corroborate the scalability experiments: even in small-scale systems where classical communication overheads are modest, ID-MAC consistently improves efficiency. As the number of QCs increases or as quantum operations become faster, its benefits scale proportionally, making it a key enabler for high-performance, large-scale quantum architectures. Moreover, by shortening the total execution time, ID-MAC indirectly enhances the system's effective coherence time, thereby improving the overall robustness and operational fidelity of multi-chip quantum systems.


\section{Conclusion}
This paper presented ID-MAC, a novel medium access control protocol designed for the classical communication subsystem of scalable multi-chip quantum architectures. The proposed approach exploits the deterministic execution flow of quantum circuits to guide token circulation dynamically, allowing the wireless channel to be accessed only by nodes with pending transmissions. By integrating compile-time knowledge of communication requirements into the MAC layer, ID-MAC eliminates idle token transitions, thereby improving channel utilization and reducing classical communication latency.

Simulation results demonstrated that ID-MAC can reduce classical communication time by up to 70\%, resulting in overall execution time reductions of 30–70\%, depending on system size and quantum technology parameters. Furthermore, by shortening execution latency, ID-MAC indirectly enhances the effective coherence time of the system, strengthening its robustness and operational reliability.

These findings highlight the crucial role of the classical communication subsystem in future large-scale quantum architectures and show that cross-layer co-design (linking circuit-level determinism with architectural control mechanisms) can significantly improve scalability and performance. Future work will focus on extending the proposed protocol to support multi-token operation.


\begingroup
\small                     
\setlength{\itemsep}{0pt}  
\setlength{\parskip}{0pt}
\linespread{0.82}\selectfont 

\bibliographystyle{IEEEtran}
\bibliography{bibliography}

\endgroup

\end{document}